\pdfoutput=1
\documentclass[conference]{IEEEtran}
\usepackage{blindtext, graphicx}
\usepackage{amsmath,amssymb}
\usepackage{bm}
\usepackage{tikz}
\usepackage{pgfplots}
\usepackage{xcolor}

\pgfplotsset{grid style={dotted,black}}
\pgfplotsset{compat=newest}

\ifCLASSINFOpdf

\else

\fi

\begin{document}
%
\title{Neural Offset Min-Sum Decoding}

\author{\IEEEauthorblockN{Loren Lugosch and Warren J. Gross}
\IEEEauthorblockA{Department of Electrical and Computer Engineering\\
McGill University
\\Email: loren.lugosch@mail.mcgill.ca, warren.gross@mcgill.ca}
}

\maketitle

\begin{abstract}
Recently, it was shown that if multiplicative weights are assigned to the edges of a Tanner graph used in belief propagation decoding, it is possible to use deep learning techniques to find values for the weights which improve the error-correction performance of the decoder. Unfortunately, this approach requires many multiplications, which are generally expensive operations. In this paper, we suggest a more hardware-friendly approach in which offset min-sum decoding is augmented with learnable offset parameters. Our method uses no multiplications and has a parameter count less than half that of the multiplicative algorithm. This both speeds up training and provides a feasible path to hardware architectures. After describing our method, we compare the performance of the two neural decoding algorithms and show that our method achieves error-correction performance within 0.1 dB of the multiplicative approach and as much as 1 dB better than traditional belief propagation for the codes under consideration.
\end{abstract}

\IEEEpeerreviewmaketitle

\section{Introduction}
Deep learning has enabled great improvement in the last few years in many inference and modelling tasks, such as speech recognition \cite{hinton2012deep}, object identification in images \cite{krizhevsky2012imagenet}, and game playing \cite{silver2016mastering}. One of the more recent applications for which deep learning has shown promise is the design of belief propagation (BP) decoders for error-correcting codes. BP is a decoding method which proceeds in iterations of message passing. In \cite{LDDL}, Nachmani \textit{et al.} showed that if the Tanner graph (the graph on which messages are passed) is ``unrolled" (that is, each iteration is considered separately) and multiplicative weights are placed on the edges of the graph, the resulting decoder is a neural network that can be trained using gradient-based methods. Such ``neural decoders" outperform traditional BP, as they learn to use their weights to mitigate the detrimental effect of cycles in the code structure. Neural decoders also require many fewer iterations and in some cases achieve a bit error rate close to that of the maximum likelihood decoder for certain high-density codes that normally decode poorly using belief propagation.

The approach presented in \cite{LDDL} requires numerous multiplications: the cost of these operations can be high and prevents efficient implementation of these decoders in complexity-critical frameworks, such as real-time hardware implementations. In this paper, we present a modification to offset min-sum decoding (an approximation to BP decoding \cite{chen2005reduced}) in which check nodes learn an additive offset parameter for each edge, rather than a multiplicative weight. Our decoding method achieves error-correction performance similar to the multiplicative weight approach in \cite{LDDL} yet uses no multiplications. It can be optimized using machine learning techniques but requires comparatively few parameters. Moreover, it incurs no extra arithmetic operations, making it fast to train.

The remainder of the paper is structured as follows. Section \ref{BP_description} describes BP and neural BP. Section \ref{NOMS_Decoding} describes the proposed technique, which is validated through a set of experiments in Section \ref{Experiments}. We compare our technique with related work in the literature in Section \ref{RelatedWork}, and we conclude with Section \ref{Conclusion}.

\section{Belief Propagation Decoding}\label{BP_description}
Before describing our contribution, we review the existing methods for belief propagation decoding of linear error-correcting codes.

\subsection{Traditional Belief Propagation}
Consider a communication system that uses a binary code and binary phase shift keying (BPSK) modulation. The transmitter sends a random vector $X \in \{-1,+1\}^n$, where $n$ is the code length. The random vector received from the communication channel is $Y = X + Z$, where $Z \in \mathbb{R}^n$ is the channel noise vector. If the channel noise is additive white Gaussian noise (AWGN), then the log-likelihood ratio (LLR) of the $v$'th received value is computed using
\begin{equation}
    l_v = \textrm{log}\left(\frac{p(Y_v = y_v | X_v = -1)}{p(Y_v = y_v | X_v = +1)}\right) = \frac{2 y_v}{\sigma^2}, 
\end{equation}
where $\sigma^2$ is the variance of the channel noise and $y$ is the observed vector of received values.

In BP decoding, probability messages are exchanged between processing nodes to compute the posterior LLRs of the received vector in an iterative fashion. At each iteration $t$, the decoder produces a ``soft output" vector $s^t$ that converges to the posterior LLRs over the iterations. The recovered binary word can be extracted from $s^t$ using a hard decision: 
\[
    \hat{x}_v^t = 
\begin{cases}
    1,& \text{if } s_v^t > 0
    \\0,              & \text{otherwise.}
\end{cases}
\]

The Tanner graph, a bipartite graph derived from the parity check matrix of the code \cite{1056404}, defines a set of variable nodes (VNs) and check nodes (CNs) used to compute the soft output. If the ($i$,$j$)-th element of the parity check matrix is a 1, then the $i$'th CN is connected to the $j$'th VN. The message $\mu_{v,c}^t$ from VN $v$ to CN $c$ for iteration $t$ is computed using

\begin{equation}\label{VC}
    \mu_{v,c}^t = l_v + \sum_{c' \in \mathcal{N}(v) \backslash c} \mu_{c',v}^{t-1},
\end{equation}
where $\mathcal{N}(v)$ is the set of CNs connected to VN $v$, and $\mu_{c,v}^0 = 0$.

The message $\mu_{c,v}^t$ from CN $c$ to VN $v$ for iteration $t$ is computed using
\begin{equation}\label{SP_CV}
    \mu_{c,v}^t = 2\textrm{tanh}^{-1}\left(\prod_{v' \in \mathcal{M}(c) \backslash v } \textrm{tanh} \left(\frac{\mu_{v',c}^t}{2}\right)\right),
\end{equation}
where $\mathcal{M}(c)$ is the set of VNs connected to CN $c$.

Finally, the soft output for iteration $t$ is computed using
\begin{equation}\label{marginalization}
    s_v^t = l_v + \sum_{c' \in \mathcal{N}(v)} \mu_{c',v}^{t}.
\end{equation}

The algorithm obtained from iteratively computing messages and outputs using Equations (\ref{VC}) to (\ref{marginalization}) is an instance of the sum-product algorithm (SPA) described in \cite{kschischang2001factor} and hence is referred to as ``SPA decoding". CNs in SPA decoding must perform multiplications and hyperbolic functions to implement Equation (\ref{SP_CV}). To simplify the CN message computation, Equation (\ref{SP_CV}) can be replaced with the ``min-sum" approximation \cite{kschischang2001factor}:
\begin{equation}\label{min_sum}
    \mu_{c,v}^t = \min_{v' \in \mathcal{M}(c) \backslash v }  (|\mu_{v',c}^t|) \prod_{v' \in \mathcal{M}(c) \backslash v } \textrm{sign} (\mu_{v',c}^t).
\end{equation}
The min-sum approximation causes non-negligible degradation in bit error rate (BER) compared to SPA. Different techniques have been proposed to overcome this issue: one of the most effective is the offset min-sum (OMS) technique \cite{chen2005reduced}, in which (\ref{min_sum}) is replaced with Equation (\ref{OMS_CV}):
\begin{equation}\label{OMS_CV}
\begin{split}
     \mu_{c,v}^t = \max\left( \min_{v' \in \mathcal{M}(c) \backslash v }  (|\mu_{v',c}^t|) - \beta, 0 \right) \\ \cdot \prod_{v' \in \mathcal{M}(c) \backslash v } \textrm{sign} (\mu_{v',c}^t),
\end{split}
\end{equation}


where $\beta$ is the correction offset. Proper selection of $\beta$ allows OMS to yield error-correction performance close to that of SPA decoding.

\subsection{Neural Belief Propagation}
If there are no cycles in the Tanner graph, then SPA exactly computes the posterior LLRs, and a hard decision on the soft output gives the optimal error-correction performance for that code. However, cycle-less codes have poor optimal performance, so codes with cycles are used in practice \cite{etzion1999codes}. BP decoding often performs well for codes with cycles, but it is not guaranteed to be optimal. High-density parity check (HDPC) codes, which have many short cycles, have poor BER when decoded using traditional SPA \cite{dimnik2009improved}.

A novel method for bringing the performance of BP closer to the optimal performance was introduced in \cite{LDDL}. In this method, the input LLRs and the messages sent between nodes are weighted using multiplicative parameters. Thus, to compute messages and outputs, Equations (\ref{VC}), (\ref{SP_CV}), and (\ref{marginalization}) are replaced with Equations (\ref{nVC}), (\ref{nSP_CV}), and (\ref{nmarginalization}), respectively:

\begin{equation}\label{nVC}
    \mu_{v,c}^t = \textrm{tanh}\left( \frac{1}{2} \left(w_{v,in}^{t} l_v + \sum_{c' \in \mathcal{N}(v) \backslash c} w_{c',v,c}^{t} \mu_{c',v}^{t-1} \right)\right)
\end{equation}

\begin{equation}\label{nSP_CV}
    \mu_{c,v}^t = 2\textrm{tanh}^{-1}\left(\prod_{v' \in \mathcal{M}(c) \backslash v } \mu_{v',c}^t\right)
\end{equation}

\begin{equation}\label{nmarginalization}
    s_v^t = w_{v,out} l_v + \sum_{c' \in \mathcal{N}(v)} w_{c',v,c,out}  \mu_{c',v}^{t},
\end{equation}



where $w_{v,in}^{t}$, $w_{c',v,c}^{t}$, $w_{v,out}$, and $w_{c',v,c,out}$ are the aforementioned multiplicative weights. Note that if these weights are equal to 1, then this approach, which we will refer to as ``neural SPA", is equivalent to SPA decoding.  While \cite{LDDL} uses SPA as the underlying BP algorithm, we have found that min-sum decoding also achieves excellent results with multiplicative weights, with minimal BER degradation compared to SPA.

A neural SPA decoder is a composition of affine functions and nonlinear functions parametrized by a set of weights, and hence can be considered a kind of neural network. Neural networks are trained using some form of gradient descent: the gradient of the ``training loss" (a measure of how poorly the network models the training data) is computed, and the gradient is used to update the weight vector. The updates continue until the training loss is minimized or acceptably low. The gradient is calculated efficiently using backpropagation, an algorithm in which the partial derivatives (informally referred to as the ``gradients") with respect to each layer of the network are calculated recursively using the chain rule. Typically, minibatch stochastic optimization is used: instead of using all training data available, an estimate of the gradient is computed using a small minibatch of examples drawn at random from the training dataset.

The work in \cite{LDDL} shows that it is possible to use minibatch stochastic gradient descent to find weights that improve the error-correction performance of a neural SPA decoder in an offline training phase. Let $w$ be the multiplicative weight vector, and let $J(w)$, the training loss to be minimized, equal
\begin{equation}
    J(w) = \mathbb{E}[H(X,S)] ,
\end{equation}
that is, the expected cross-entropy between the decoder output $S$ and the transmitted codeword in binary format $X$. The cross-entropy, $H(X,S)$, is defined as 
\begin{equation}
    H(X,S) = -\frac{1}{n}\sum_v  X_v \textrm{log} (\sigma(S_v)) + (1-X_v) \textrm{log} (1 - \sigma(S_v)) ,
\end{equation}
where $\sigma()$, the sigmoid nonlinearity, is a soft, differentiable version of the hard decision,
\begin{equation}
    \sigma(x) = \frac{1}{1+e^{-x}}.
\end{equation}

The value of $J(w)$ for a particular $w$ can be estimated by computing the average cross-entropy across a minibatch of training codewords.  Since every operation in neural SPA is differentiable, it is possible to compute the gradient of $J$ with respect to the parameters, $\nabla_w J$, using backpropagation. The parameters can then be incrementally updated using the gradient descent update rule or a variant thereof \cite{ngiam2011optimization}.

Since CN and VN operations in neural SPA satisfy the message passing symmetry conditions in \cite{richardson2008modern}, the probability of error is independent of the transmitted codeword. Thus, training data can be formed by adding AWGN to the all-zeros codeword.

\section{Neural Offset Min-Sum Decoding}\label{NOMS_Decoding}
In this section, we introduce neural offset min-sum (NOMS) decoding, a learning algorithm which achieves performance similar to that of neural SPA but is more efficient and potentially easier to implement in hardware. NOMS is a generalization of OMS; unlike OMS, which uses a single global offset, NOMS uses multiple learnable offset parameters. The learnable offsets serve the dual purpose of reducing error caused by the min-sum approximation and attenuating cycles in the Tanner graph. In this section, we describe how decoding and training proceed in NOMS.

\subsection{Decoding}
Let ReLU(), the ``rectifier" activation function commonly used in deep learning \cite{nair2010rectified}, be defined as
\begin{equation}
    \textrm{ReLU}(x) = \max(x,0).
\end{equation}
Thus, it is possible to rewrite Equation (\ref{OMS_CV}) more compactly as 
\begin{equation}\label{rOMS_CV}
    \mu_{c,v}^t = \textrm{ReLU}\left( \min_{v'}  (|\mu_{v',c}^t|) - \beta \right) \prod_{v'} \textrm{sign} (\mu_{v',c}^t)
\end{equation}
for $v' \in \mathcal{M}(c) \backslash v$. 

Messages between VNs and CNs in NOMS are computed using the same equations as OMS, except Equation (\ref{rOMS_CV}) is replaced with
\begin{equation}\label{NOMS_CV}
    \mu_{c,v}^t = \textrm{ReLU}\left( \min_{v'}  (|\mu_{v',c}^t|) - \beta_{c, v}^t \right) \prod_{v'} \textrm{sign} (\mu_{v',c}^t),
\end{equation}
where $\beta_{c,v}^t$ is a learnable offset parameter for the edge connecting CN $c$ to VN $v$ during iteration $t$. As is the case for neural SPA decoders, the error-correction performance of a NOMS decoder is independent of the transmitted codeword.

\subsection{Training}
Since all of the operations used in NOMS decoding are differentiable, the offset parameters can be learned using minibatch stochastic gradient descent in an offline phase, as described in Section \ref{BP_description}. A potential problem for backpropagation arises in that some of the constituent operations (ReLU(), min(), abs(), and sign()) are not differentiable at certain points. Deep learning frameworks such as TensorFlow overcome this problem by simply choosing one of the subgradients at the points of non-differentiability in order to backpropagate through piecewise differentiable operations \cite{tensorflow2015}. This is the approach we take to train NOMS decoders. (Note that since the derivative of sign() is equal to 0 almost everywhere, gradients in the NOMS decoder effectively only flow through the ReLU() branch of the CN.)

\subsection{Implementation Cost}
NOMS can achieve good performance while using substantially fewer operations and parameters than the version of neural SPA presented above. For an $(n,k)$ code with $E$ edges in the Tanner graph and check nodes of degree $d_c$, with $T$ iterations of BP unrolled, neural SPA requires $nT + ET(d_c-1) + n + nE$ parameters, whereas our approach requires only $ET$ parameters. Likewise, decoding a received vector using neural SPA requires $nT + ET(d_c-1) + n + nE$ additional multiplications, while our approach, like OMS, requires $ET$ additions compared to pure min-sum decoding (and therefore no additional arithmetic operations compared to OMS).

\subsection{Constraining Offset Count}
While we only present results in this paper for the case when each edge has its own learnable offset, the number of offsets used by the decoder can be reduced by constraining multiple edges to using the same offset. In particular, if the decoder is constrained to use a single offset for all edges, then NOMS reverts to OMS. Indeed, constraining the decoder to learning a single offset yields a new way to choose the offset for OMS, in addition to the existing methods of simulation and density evolution.

\section{Experiments}\label{Experiments}
\subsection{Experimental Setup}
We compared the performance of NOMS decoding with both traditional SPA and neural SPA, unrolled to 5 iterations. The codes we tested were the same codes as were used in \cite{LDDL}, namely, BCH(63,36), BCH(63,45), and BCH(127,106). BCH codes typically do not decode well using traditional BP algorithms, and are therefore a good test case for highlighting the improvement due to learnable parameters. Note that we reduced the number of weights in the neural SPA decoder for the BCH(127,106) code.

To evaluate the performance of the decoders, we performed Monte-Carlo simulation. We created test data by generating random binary messages, encoding them using the BCH generator matrices, modulating the codewords using BPSK, and adding AWGN channel noise. To minimize the variance of the BER estimates, we required a minimum of 100 frame errors to be detected and 100,000 frames to be simulated for each signal-to-noise ratio (SNR).

To train the decoders, we used 20,000 minibatches of 120 received words drawn uniformly from each SNR to simulate and using only the all-zeros codeword as the transmitted codeword. We initialized the offsets to random values drawn from the standard normal distribution. To update the offsets after calculating the gradients, we used the Adam optimizer \cite{ADAM} with a learning rate of 0.1.

\subsection{Results}
\begin{figure}
    \centering
    \begin{tikzpicture}
        \begin{semilogyaxis}[
            height=9cm,
            width=9cm,
            grid=both,
            xlabel=$E_b / N_0$,
            ylabel=BER
        ]
    
        \addplot coordinates {
            (1,0.11348079365079365)
            (2,0.0816704761904762)
            (3,0.049108095238095235)
            (4,0.024254603174603175)
            (5,0.01030047619047619)
            (6,0.00344)
            (7,0.0010253968253968254)
            (8,0.0001995238095238095)
        };
        \addlegendentry{SPA}
    
        \addplot coordinates {
            (1, 1.13834656e-01 )
            (2,   8.24378307e-02 )
            (3,   4.66917989e-02 )
            (4,   1.89285714e-02 )
            (5, 5.00529101e-03   )
            (6, 8.21957672e-04   )
            (7, 7.65873016e-05   )
            (8, 5.15873016e-06)
        };
        \addlegendentry{[4]}
        
        \addplot coordinates {
            (1,0.11493218123913568)
            (2,0.08690063821958306)
            (3,0.05621931026607285)
            (4,0.026479134152995066)
            (5,0.00740027025998249)
            (6,0.0012407534290028294)
            (7,0.00015622422696763224)
            (8,1.4258045207885122e-05)
        };
        \addlegendentry{NOMS}
        \end{semilogyaxis}
    \end{tikzpicture}
    \caption{BER for BCH (63,36) code.}
    \label{fig:BCH_63_36}
\end{figure}
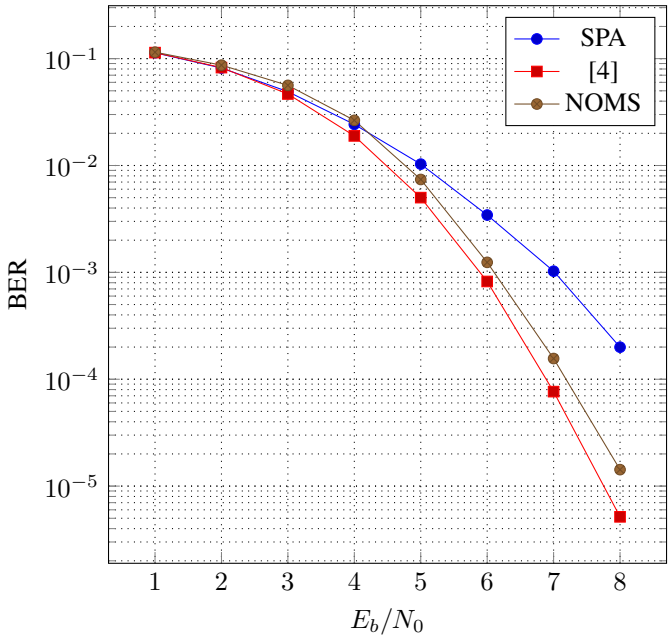

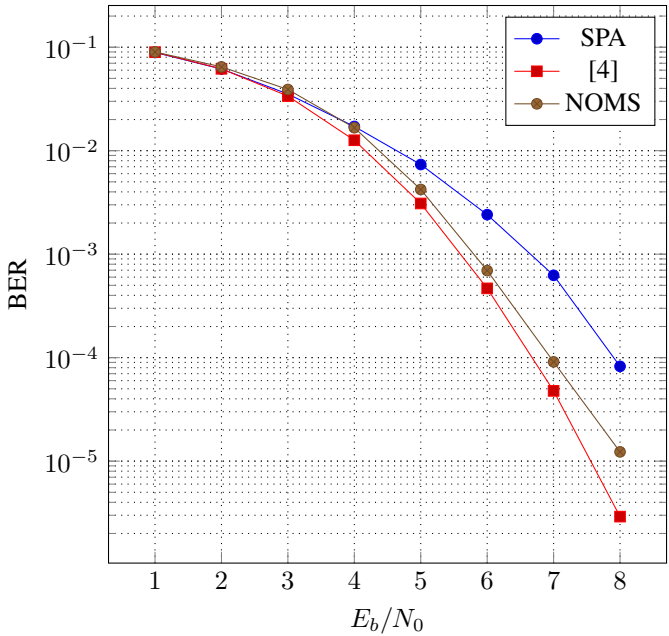
\begin{figure}
    \centering
\begin{tikzpicture}
        \begin{semilogyaxis}[
            height=9cm,
            width=9cm,
            grid=both,
            xlabel=$E_b / N_0$,
            ylabel=BER
        ]

        \addplot coordinates {
            (1,8.89904797e-02)
            (2,6.15665093e-02)
            (3,3.53112705e-02)
            (4,1.71428565e-02)
            (5,7.35365087e-03)
            (6,2.41063489e-03)
            (7,6.24761917e-04)
            (8,8.24175804e-05)
        };
        \addlegendentry{SPA}
    
        \addplot coordinates {
            (1, 8.95687831e-02)
            (2, 6.18478836e-02)
            (3, 3.37261905e-02)
            (4, 1.26177249e-02)
            (5, 3.08994709e-03)
            (6, 4.67328042e-04)
            (7, 4.77513228e-05)
            (8, 2.91005291e-06)
        };
        \addlegendentry{[4]}
        
        \addplot coordinates {
            (1,0.0899547029043432)
            (2,0.06459990737568676)
            (3,0.039068427797444585)
            (4,0.01667285219443493)
            (5,0.00421805412812607)
            (6,0.0006943651428063898)
            (7,9.103828048672173e-05)
            (8,1.2290081026748483e-05)
        };
        \addlegendentry{NOMS}
        \end{semilogyaxis}
    \end{tikzpicture}
    \caption{BER for BCH (63,45) code.}
    \label{fig:BCH_63_45}
\end{figure}

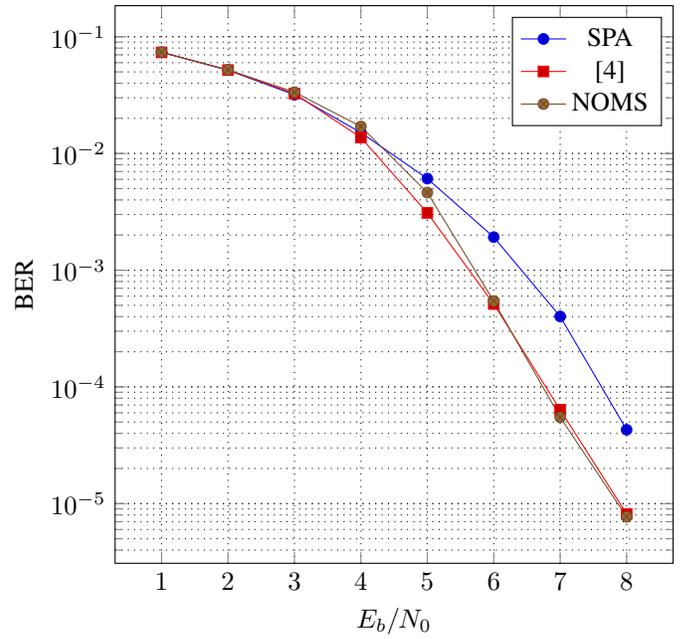
\begin{figure}
    \centering
\begin{tikzpicture}
        \begin{semilogyaxis}[
            height=9cm,
            width=9cm,
            grid=both,
            xlabel=$E_b / N_0$,
            ylabel=BER
        ]

        \addplot coordinates {
            (1,0.07360984251968504)
            (2,0.051766771653543306)
            (3,0.03184834645669291)
            (4,0.015090629921259842)
            (5,0.006090708661417323)
            (6,0.001921023622047244)
            (7,0.000401496062992126)
            (8,4.292432195975503e-05)
        };
        \addlegendentry{SPA}

        \addplot coordinates {
            (1,0.07345816260377525)
            (2,0.051815791461319136)
            (3,0.032605883796899486)
            (4,0.013695185583816412)
            (5,0.0030964205013941605)
            (6,0.0005131328008459374)
            (7,6.364986750756875e-05)
            (8,8.16391096482831e-06)
        };
        \addlegendentry{[4]}

        \addplot coordinates {
            (1,0.07379238341610177)
            (2,0.05209934414672986)
            (3,0.03354497504358718)
            (4,0.017030470112099298)
            (5,0.004622522454477363)
            (6,0.0005438169149719594)
            (7,5.4916696564008636e-05)
            (8,7.707291615798542e-06)
        };
        \addlegendentry{NOMS}
        \end{semilogyaxis}
    \end{tikzpicture}
    \caption{BER for BCH (127,106) code.}
    \label{fig:BCH_127_106}
\end{figure}

Fig. \ref{fig:BCH_63_36}-\ref{fig:BCH_127_106} show the performance of the different decoding methods. At low SNRs, both neural decoding methods have roughly the same BER as SPA decoding. At higher SNRs, the neural decoders outperform SPA by as much as 1 dB. NOMS performs close to neural SPA until the highest SNRs simulated.

\begin{figure}
    \centering

    \begin{tikzpicture}
    \begin{axis}[
        yticklabel style={
            /pgf/number format/fixed,
            /pgf/number format/fixed zerofill,
            /pgf/number format/precision=0
        },
        ybar=0pt, 
        bar width=0.2cm,
        height=9cm,
        width=9cm,
        xlabel=Offset value,
        ylabel=Frequency
    ]
    \addplot [fill=green, fill opacity=1] table [x, y, col sep=comma]{iteration_1_100.csv}; \addlegendentry{Iteration 1}
    \addplot [fill=red, fill opacity=1] table [x, y, col sep=comma]{iteration_2_100.csv}; \addlegendentry{Iteration 2}
    \addplot [fill=blue, fill opacity=1] table [x, y, col sep=comma]{iteration_3_100.csv}; \addlegendentry{Iteration 3}
    \addplot [fill=yellow, fill opacity=1] table [x, y, col sep=comma]{iteration_4_100.csv}; \addlegendentry{Iteration 4}
    \addplot [fill=violet, fill opacity=1] table [x, y, col sep=comma]{iteration_5_100.csv}; \addlegendentry{Iteration 5}
    \end{axis}
    \end{tikzpicture}

    \caption{Offset histogram for BCH (63,45) decoder after 100 minibatches.}
    \label{fig:histogram_63_45_1}
\end{figure}
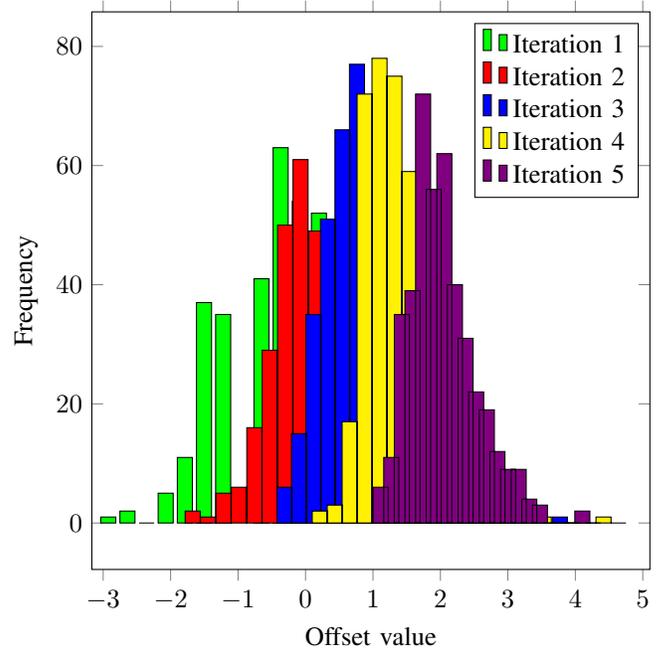

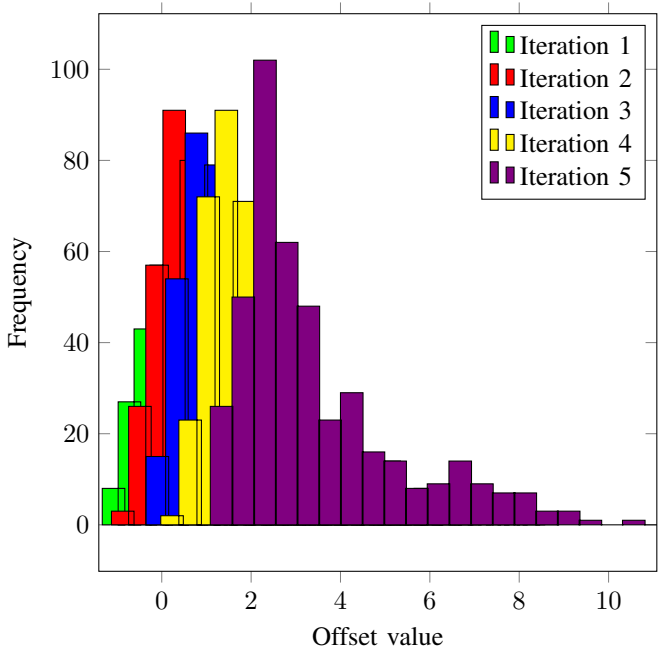
\begin{figure}
    \centering
    \begin{tikzpicture}
    \begin{axis}[
        yticklabel style={
            /pgf/number format/fixed,
            /pgf/number format/fixed zerofill,
            /pgf/number format/precision=0
        },
        ybar=0pt, 
        bar width=0.3cm,
        height=9cm,
        width=9cm,
        xlabel=Offset value,
        ylabel=Frequency
    ]
    \addplot [fill=green, fill opacity=1] table [x, y, col sep=comma]{iteration_1_3000.csv};
    \addlegendentry{Iteration 1}
    \addplot [fill=red, fill opacity=1] table [x, y, col sep=comma]{iteration_2_3000.csv}; \addlegendentry{Iteration 2}
    \addplot [fill=blue, fill opacity=1] table [x, y, col sep=comma]{iteration_3_3000.csv}; \addlegendentry{Iteration 3}
    \addplot [fill=yellow, fill opacity=1] table [x, y, col sep=comma]{iteration_4_3000.csv}; \addlegendentry{Iteration 4}
    \addplot [fill=violet, fill opacity=1] table [x, y, col sep=comma]{iteration_5_3000.csv}; \addlegendentry{Iteration 5}
    \end{axis}
    \end{tikzpicture}
    
    \caption{Offset histogram for BCH (63,45) decoder after 3,000 minibatches.}
    \label{fig:histogram_63_45_2}
\end{figure}

Another noteworthy aspect of the decoder is the offset distribution. Fig. \ref{fig:histogram_63_45_1} and Fig. \ref{fig:histogram_63_45_2} show 20-bin histograms of the offsets for different iterations after the decoder has been trained on 100 minibatches and 3,000 minibatches, respectively. Early in the training process (Fig. \ref{fig:histogram_63_45_1}), the range of values which the offsets are shown to take on remains for the most part close to the standard normal distribution to which all offsets are initialized. Offsets for the third, fourth, and fifth iterations take on values consistent with common values for $\beta$ reported in the OMS literature (see e.g. \cite{7378848}), suggesting that early in training, the network has at least partly learned to correct the min-sum approximation. Later (Fig. \ref{fig:histogram_63_45_2}), a second mode appears in the histogram for each iteration in a much higher range (at a value of around 7). This occurs when the BER curve of the NOMS decoder starts to match the BER curve of the neural SPA decoder, meaning that after this much training, the NOMS decoder is now no longer only correcting the min-sum approximation, but has also discovered a way to attenuate cycles. 

\section{Related Work}\label{RelatedWork}
In \cite{zhang2006two}, Zhang \textit{et al.} note that for irregular LDPC codes, OMS can incur a significant BER degradation compared to SPA decoding. To fight this effect, they suggest having each CN of degree $d$ use offset $\beta_d$, rather than a global $\beta$. The authors use a genetic algorithm to choose the offsets for each degree and, interestingly, find that this offsetting scheme not only matches the performance of SPA but can also lower the BER error floor exhibited by some codes. 

The approach given in \cite{zhang2006two} is similar to NOMS in that it uses more than just one global offset. The main difference between this approach and NOMS is that the offsets found by \cite{zhang2006two} do not depend on particular paths through the Tanner graph, and thus in general may have difficulty overcoming the effect of short cycles, especially if the code has only a few different check node degrees. For the BCH codes considered in this work, for example, every check node has the same degree, so the approach in \cite{zhang2006two} would yield a single $\beta_d$, thus reverting to OMS. NOMS, on the contrary, has the potential to improve performance even for regular codes. Finally, the genetic algorithm used to learn the offsets in \cite{zhang2006two} is a random heuristic method, which, unlike gradient-based optimization methods, has been found to perform poorly as the size of the optimization problem increases \cite{rios2013derivative}. 

\section{Conclusion}\label{Conclusion}
In this work, we proposed NOMS, a new decoding algorithm which generalizes OMS decoding using additive offsets optimized by treating the decoder as a deep neural network. NOMS grants substantial error-correction performance improvements (up to 1 dB for the codes considered here) at a small cost in complexity. In the future, we plan to explore the effect of constraining the number of offsets on BER compared to unconstrained NOMS. We also plan to test NOMS on various other types of codes, such as LDPC codes.

\section*{Acknowledgment}
The authors would like to thank NVIDIA's Academic Grant Program for providing the GPU we used for training and testing the decoders. We would also like to thank Dr. Carlo Condo for his helpful comments in the preparation of this paper.

\ifCLASSOPTIONcaptionsoff
  \newpage
\fi

\bibliographystyle{IEEEtran}
\bibliography{references}

\end{document}